\pdfoutput=1
\documentclass[twoside,english,sort&compress]{iopart}
\usepackage[T1]{fontenc}
\usepackage[latin9]{inputenc}
\usepackage{geometry}
\usepackage{float}
\usepackage{bm}
\usepackage{graphicx}
\usepackage[numbers]{natbib}
\usepackage{color}
\usepackage{lineno}

\makeatletter

\usepackage{iopams}
\usepackage{setstack}

\newcommand{\eqref}[1]{(\ref{#1})}

\makeatother

\usepackage{babel}
\begin{document}

\title{Dynamics of cold pulses induced by super-sonic molecular beam injection in the EAST tokamak}

\author{Yong Liu$^1$, Yuejiang Shi$^2$, Tao Zhang$^1$, Chu Zhou$^3$, Xiaolan Zou$^4$, Hailin Zhao$^1$, Ahdi Liu$^3$, Tianfu Zhou$^1$, Xiang Liu$^1$, Shoubiao Zhang$^1$, Bin Cao$^1$, and Volker Naulin$^5$}

\address{$^1$ Institute of Plasma Physics, Chinese Academy of Sciences, Hefei 230031, People's Republic of China \\
$^2$ Department of Nuclear Engineering, Seoul National University, Seoul, Republic of Korea \\
$^3$ University of Science and Technology of China, Hefei 230026, China \\
$^4$ CEA, IRFM, F-13108 Saint-Paul-lez-Durance, France\\
$^5$ Technical University of Denmark (DTU), Department of Physics, DK-2800 Kgs. Lyngby, Denmark
}

\ead{zhouchu@ustc.edu.cn}

%\linenumbers

\begin{abstract}
Evolution of electron temperature, electron density and its fluctuation with high spatial and temporal resolutions are presented for the cold pulse propagation induced by super-sonic molecular beam injection (SMBI) in ohmic plasmas in the EAST tokamak. The non-local heat transport occurs for discharges with plasma current $I_p$=450 kA ($q_{95}\sim5.55$), and electron density $n_{e0}$ below a critical value of $(1.35\pm0.25)\times10^{19}~\mathrm{m^{-3}}$. In contrary to the response of core electron temperature and electron density (roughly 10 ms after SMBI), the electron density fluctuation in the plasma core increases promptly after SMBI and reaches its maximum around 15 ms after SMBI. The electron density fluctuation in the plasma core begins to decrease before the core electron temperature reaches its maximum (roughly 30 ms). It was also observed that the turbulence perpendicular velocity close to the inversion point of the temperature perturbation changes sign after SMBI.
\end{abstract}
%\maketitle

One of the most provocative phenomenons in plasma transport is the so called non-local heat transport: the core temperature increases in response to edge cooling. This was observed for the first time in the TEXT experiment in 1995 \cite{gentle1995}, and was reproduced in many tokamaks (TFTR \cite{kissick1996}, Tore Supra \cite{zou2000}, RTP \cite{mantica1999}, ASDEX-U \cite{ryter2000}, JET \cite{mantica2002}, HL-2A \cite{sun2010}, Alcator C-Mod \cite{rice2013}, KSTAR \cite{shi2017}, J-TEXT \cite{shi2018}) and helical devices (LHD \cite{inagaki2006}). Recent results in low-current Ohmic plasmas in Alcator C-Mod \cite{rice2013,gao2014} and in ECH-heated plasmas in KSTAR \cite{shi2017} indicate that the critical collisionality are closely related for the appearance of the non-local heat transport and the intrinsic toroidal rotation reversal, and this suggests a connection between heat and momentum transport. This may be explained by a model with a transition of dominant turbulence from trapped electron modes (TEMs) to  ion temperature gradient (ITG) modes when the collisionality increases. However, latest results in Alcator C-Mod \cite{fernandez2017} show that 
the disappearance of the non-local heat transport and the intrinsic toroidal rotation reversal are not concomitant in high-current Ohmic plasmas
and ICRH-heated L-mode plasmas. This suggests that the heat transport and the momentum transport are driven by different micro-instabilities under certain circumstances. 

Presently, to the authors' knowledge, there is no well-accepted theories/models to interpret this phenomenon. In the early times date back to late 1990s \cite{callen1997}, various types of empirical model, marginal-stability-based models and self-organized criticality based models, have been proposed to interpret the observations. Later on, other models, including fractional diffusion model \cite{negrete2008} and turbulence spreading transport model \cite{hariri2016}, have been developed. Very recent simulation results \cite{fernandez2018} using the newly developed trapped gyro-Landau fluid model TGLF-SAT1 have reproduced the cold-pulse dynamics at Alcator C-Mod. It was found that the cold-pulse phenomena can be explained by the competition between density gradient driven TEM and ITG turbulence. However, the evolution of the electron density profiles is not available experimentally in Alcator C-Mod with high temporal resolution. Therefore the  electron density profiles used in the simulation \cite{fernandez2018} are constructed by introducing an inwardly propagating skewed Gaussian, using the experimental constraints of the measured total radiated power and line-averaged density.

\begin{figure}[b]
\centering
\caption{\label{fig:los}General waveforms for a discharge the non-local transport occurs (Left, the red lines indicate the SMBI events): (a) plasma current, (b) core electron density, (c) core electron temperature, and count rates of $\gamma$ ray (0.3-6 MeV) (d) and hard X ray (20-200 keV) (e), and the line of sight for the essential diagnostics (Right) used in this study (The blue solid line is the ray with frequency of 70 GHz).}
\includegraphics[width=6cm]{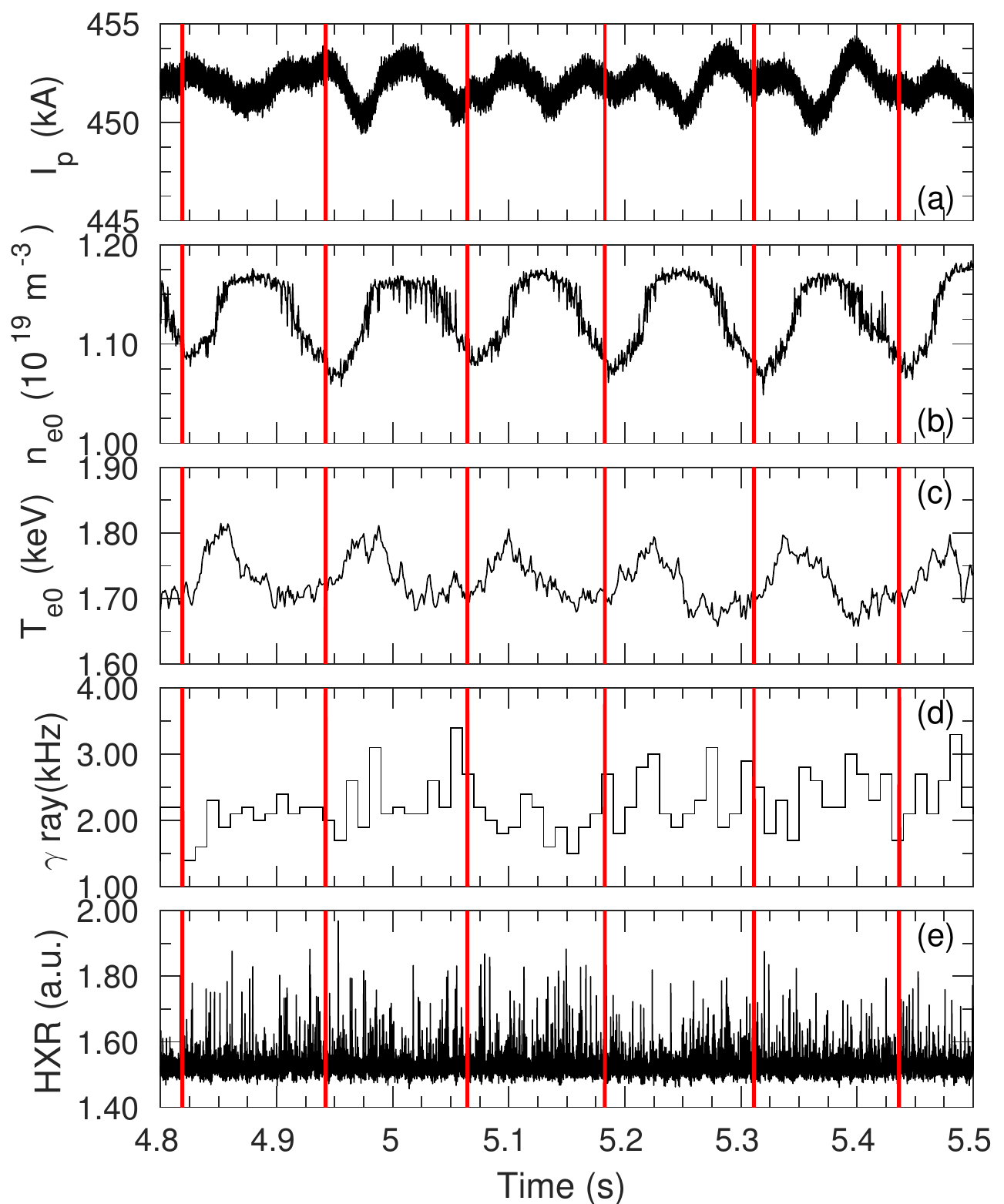}
\includegraphics[width=3.8cm]{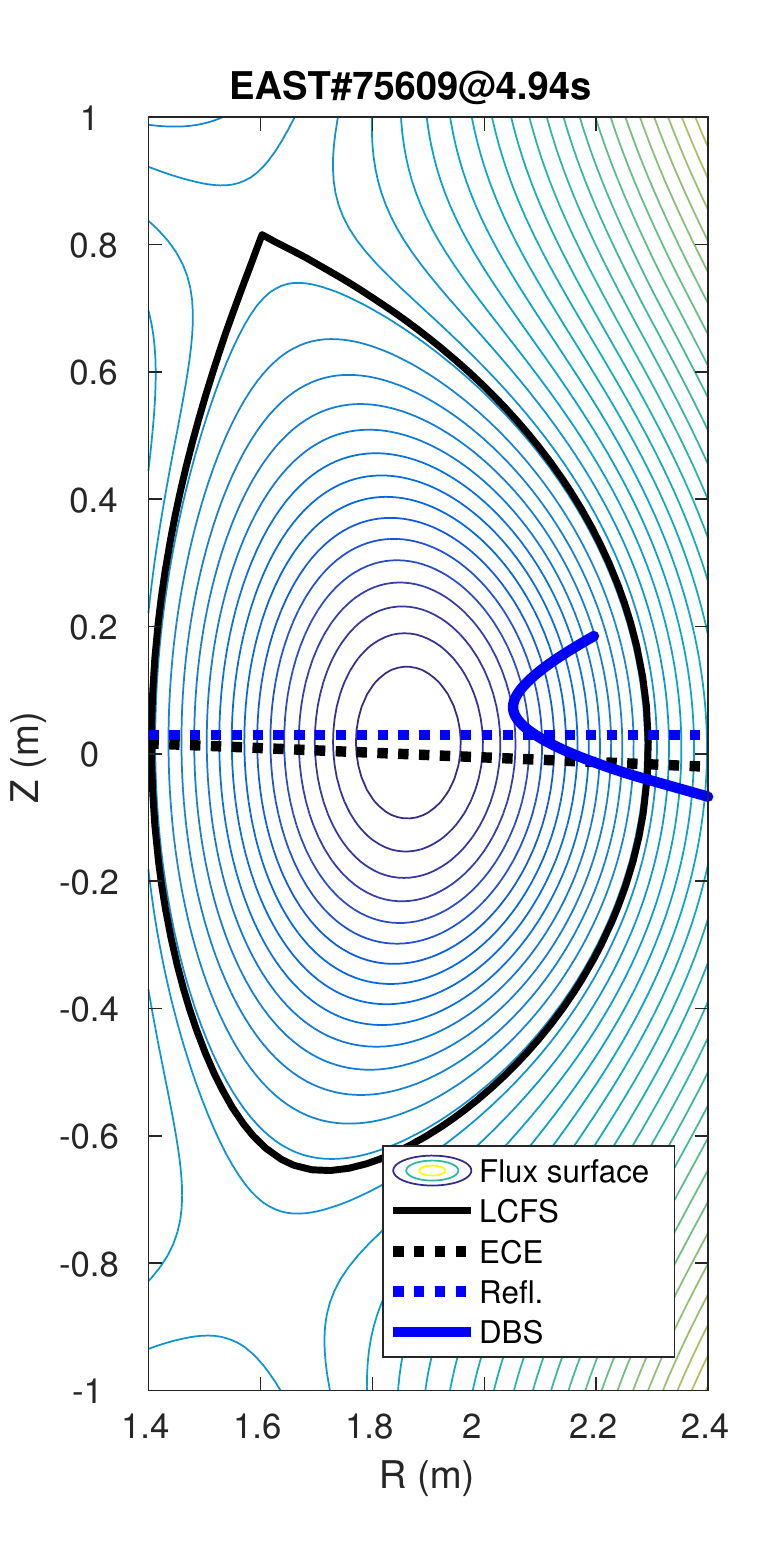}
\end{figure}

In EAST non-local heat transport has been observed recently during plasma edge cooling induced by super-sonic molecular beam injection (SMBI). The discharges discussed in this work are ohmic plasmas with upper single null configuration. The main plasma parameters are: plasma current $I_p$=450 kA, magnetic axis $R_m$=1.87 m, minor radius $a$=0.45 m, toroidal magnetic field $B_t$=2.24 T. The time interval between SMBI pulses roughly varies from 120 ms to 150 ms, and the SMBI pulse duration is 1 ms. During an SMBI pulse 1.6$\times10^{19}$ molecules are injected into the tokamak. The electron temperature is measured by an electron cyclotron emission (ECE) radiometer system \cite{liu2014} with a temporal resolution of up to 2.5 $\mu$s. The electron density is measured by a microwave reflectometry \cite{qu2015} with a temporal resolution of up to 0.5 ms. The electron density fluctuation and its perpendicular velocity are simultaneously measured by a multi-channel Doppler backscattering (DBS) system \cite{zhou2013}. Figure \ref{fig:los} shows the general waveforms of a discharge where the non-local heat transport occurs (Left): (a) plasma current, (b) core electron density, (c) core electron temperature, and count rates of $\gamma$ ray (0.3-6 MeV) (d) and hard X ray (20-200 keV) (e), and the line of sights for the essential diagnostics used in this study (Right). Clearly both the core electron density and the electron temperature increase in response to SMBI. For the edge we observe the density increase and cooling as expected to be induced by SMBI.  No correlation between electron temperature and supra-thermal electrons (indicated by the count rates of $\gamma$ ray and hard X ray) is observed.

\begin{figure}[htbp]
\centering
\caption{\label{fig:waveforms75609}Evolution of electron temperature and electron density: (a) Electron temperature at major radius R=2.2 m, (b) Electron temperature at R=1.95 m, (c) Electron density at R=2.2 m, (d) Electron density at R=1.95 m. The red lines indicate the SMBI events.}
\includegraphics[width=8cm]{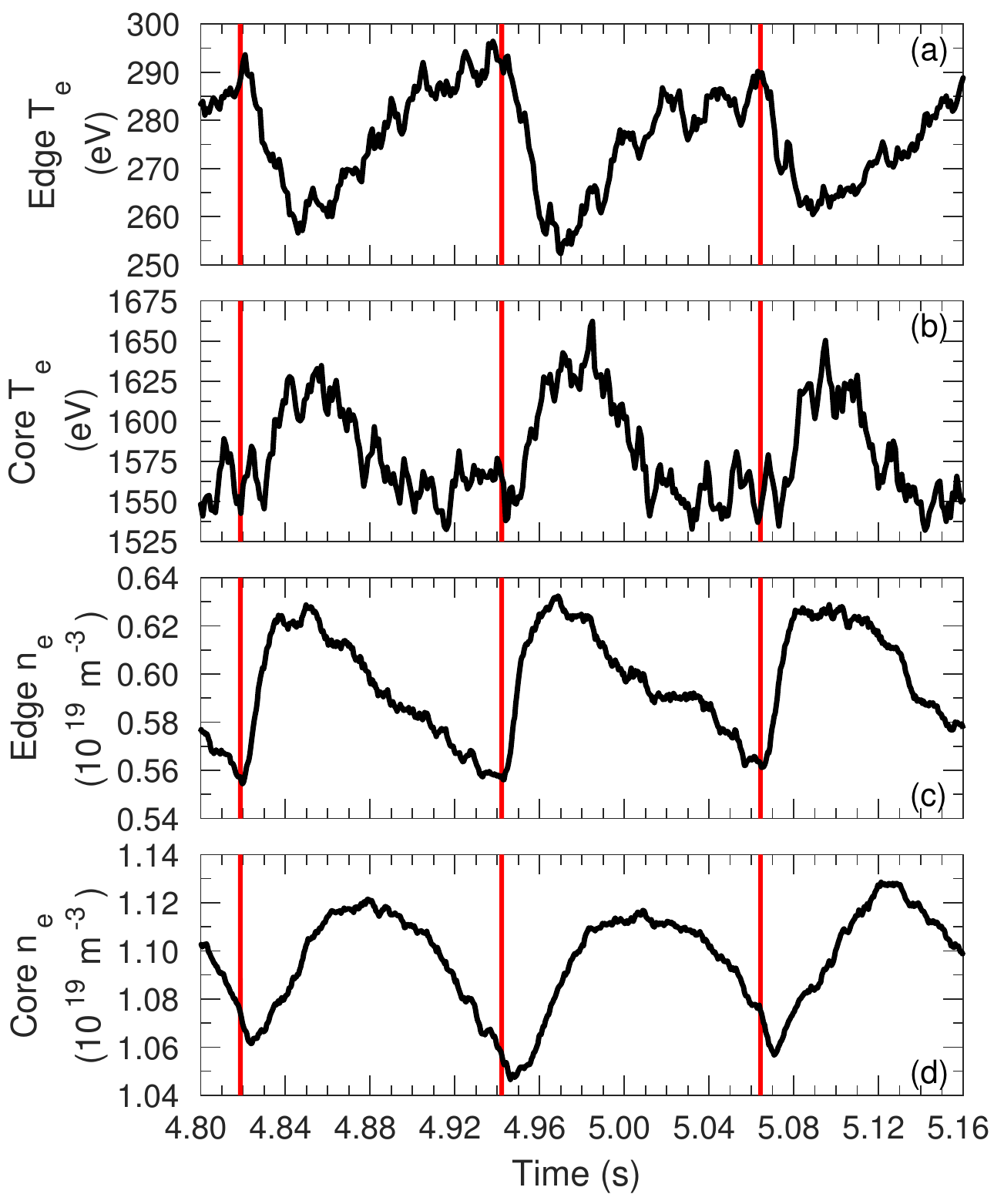}
\end{figure}

Figure \ref{fig:waveforms75609} shows the evolution of electron temperature and electron density respectively at plasma periphery and core. The electron density $n_{e0}$ close to the magnetic axis is roughly $1.1\times10^{19}~\mathrm{m^{-3}}$, and increases by less than 10\% induced by SMBI (see Figure \ref{fig:waveforms75609} (d)). Both the electron temperature and the electron density at the plasma periphery have a prompt response to the SMBIs. A clear increase of the core electron temperature is  observed, and a delay of a few ms exists. There also exists a delay in the response of core electron density. A discharge (EAST\#75592) with similar parameters but higher electron density ($n_{e0}\approx1.6\times10^{19}~\mathrm{m^{-3}}$) shows standard transport behavior. The critical cutoff density is determined to be $(1.35\pm0.25)\times10^{19}~\mathrm{m^{-3}}$ for $I_p$=450 kA ($q_{95}\sim5.55$), and this is in a good agreement with the scaling law proposed in the reference \cite{gao2014}.

\begin{figure}[htbp]
\flushright
\caption{\label{fig:a_smbi75609}Evolution (relative to SMBI) of electron temperature (Left), electron density (Middle), and normalized electron density gradient $a/L_{n}$ (Right) at different radii relative to the SMBI event at 4.942 s in Figure \ref{fig:waveforms75609}.}
\includegraphics[width=12cm]{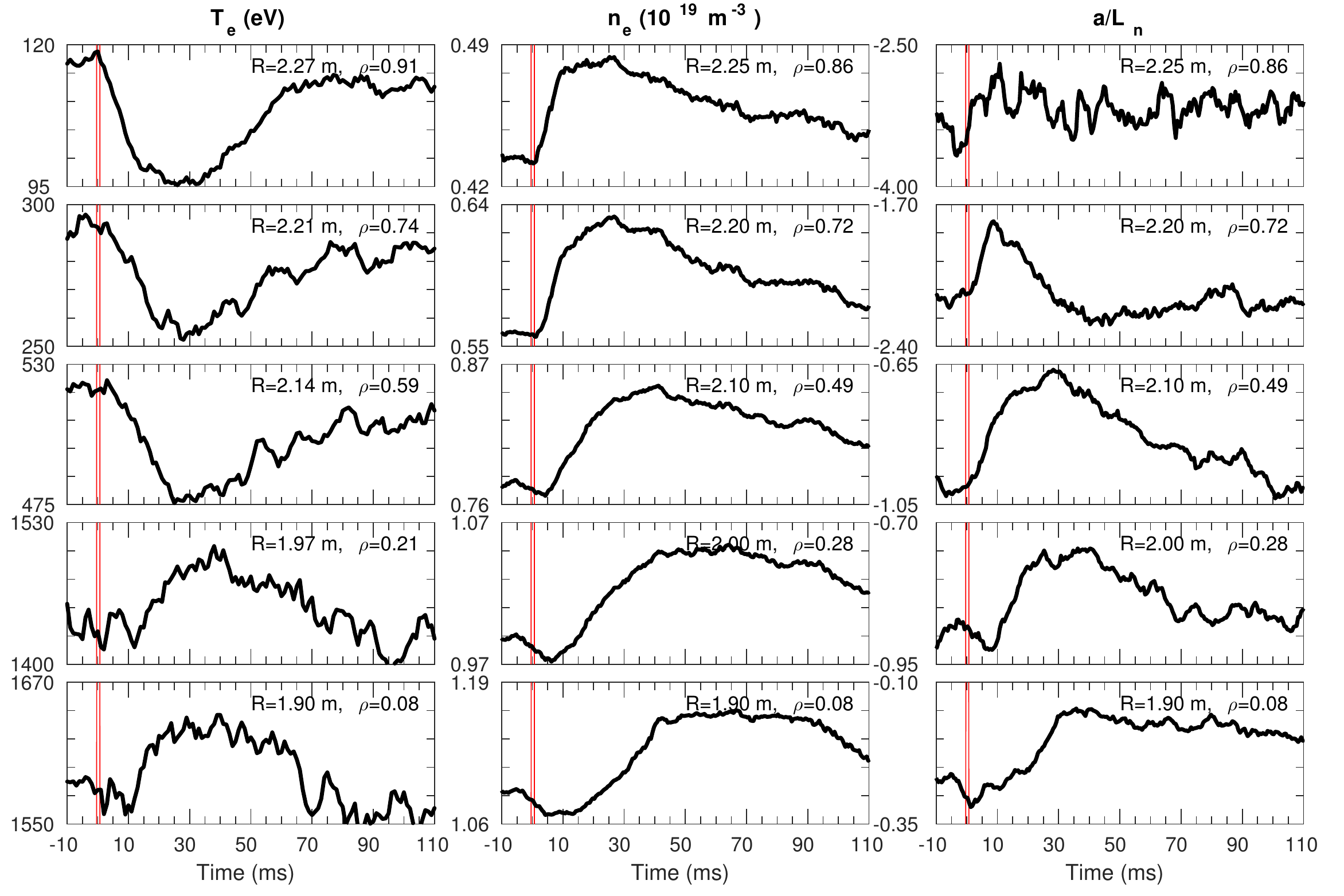}
\end{figure}

\begin{figure}[htbp]
\centering
\caption{\label{fig:a_smbi_2DTene75609}Evolution of the normalized electron temperature and density. They are respectively defined as $T_e/T_{e0}-1$ and $(n_e-n_{e0})/(n_{emax}-n_{e0})$, where $T_{e0}$ and $n_{e0}$ are the electron temperature and density at the time SMBI is on, $n_{emax}$ is the maximum electron density after a SMBI event.}
\includegraphics[width=8cm]{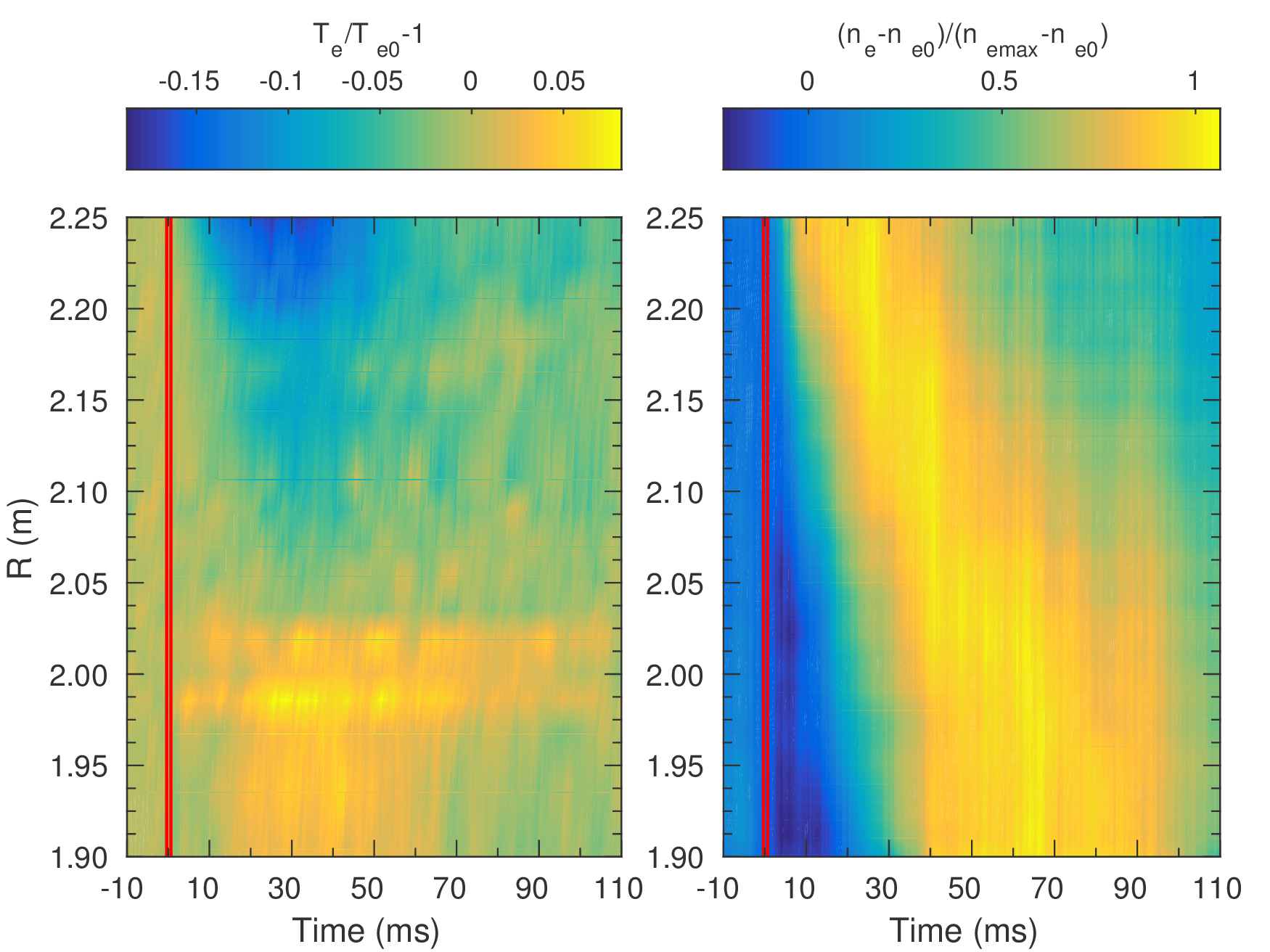}
\end{figure}

Evolution (relative to SMBI) of electron temperature, electron density, and normalized electron density gradient $a/L_{n}$ at different radii for the SMBI event at time 4.942 s are illustrated in Figure \ref{fig:a_smbi75609}. The electron temperature and density in plasma core begin to increase roughly 10 ms after SMBI. It is not obvious to see the cold pulse propagation from the evolution of electron temperature, and the core electron temperature reaches its maximum almost at the same time when the edge electron temperature reaches its minimum. It is clear that the time for the electron density to reach its maximum becomes larger from plasma edge to core, and this is well depicted in Figure \ref{fig:a_smbi_2DTene75609} that shows the evolution of the normalized electron density ($(n_e-n_{e0})/(n_{emax}-n_{e0})$ ($n_{e0}$ is the electron density at the time SMBI is on, and $n_{emax}$ is the maximum electron density after a SMBI event). The results clearly show inwardly propagation of the density perturbation induced by SMBI. Consequently, the inwardly propagation of density perturbation results in the variation of the normalized electron density gradient $a/L_n$. It is clear and easy to understand that the local $a/L_n$ reaches its minimum slightly before the local electron density reaches its maximum.

\begin{figure}[htbp]
\centering
\caption{\label{fig:dbs_amp_75609}Evolution of the normalized amplitude of the scattering signals measured by DBS system. }
\includegraphics[width=8cm]{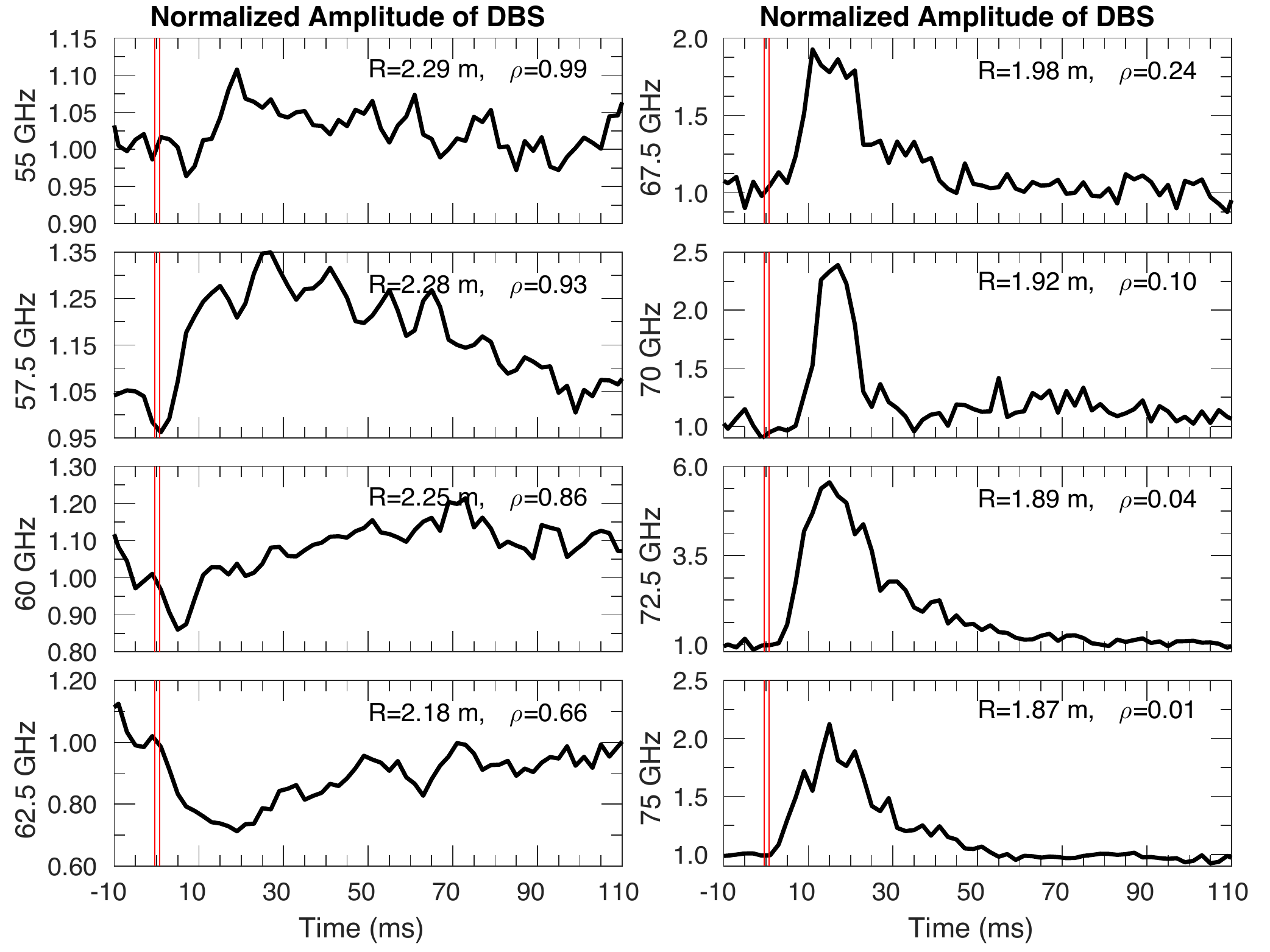}
\end{figure}

The localized electron density fluctuations can be measured by DBS in EAST. Figure \ref{fig:dbs_amp_75609} shows the evolution of the normalized amplitude (proportional to the electron density fluctuation) of the scattering signals measured by DBS system. As the results show, the electron density fluctuation in the plasma edge (the two channels with lowest probing frequencies) increases slightly. This is a common phenomenon accompanying with SMBI on EAST. In the plasma core region ($\rho$<0.3), the electron density fluctuation increases remarkably. This is different to the observation on Alcator C-Mod \cite{gao2014}. It was found from phase contrast imaging diagnostic that the density fluctuations are suppressed during the laser blow-off impurity injections for the Alcator C-Mod discharge where the non-local effect of edge cooling exists. It is spectacular that the electron density fluctuation in the plasma core responses to SMBI promptly. This coincides with the behavior observed in simulations with the turbulence spreading model \cite{hariri2016}. More spectacularly, the electron density fluctuation decreases significantly at the location $\rho=0.66$.

\begin{figure}[htbp]
\flushright
\caption{\label{fig:u_perp75609} (Left) Radial profiles of velocity ($u_\perp$) extracted from the Doppler shift measured by DBS system. The cross indicates error bar of the measurement, and error bars for other data points are more or less the same and are not shown. (Right) Electron temperature profiles before and after SMBI.}
\includegraphics[width=6cm]{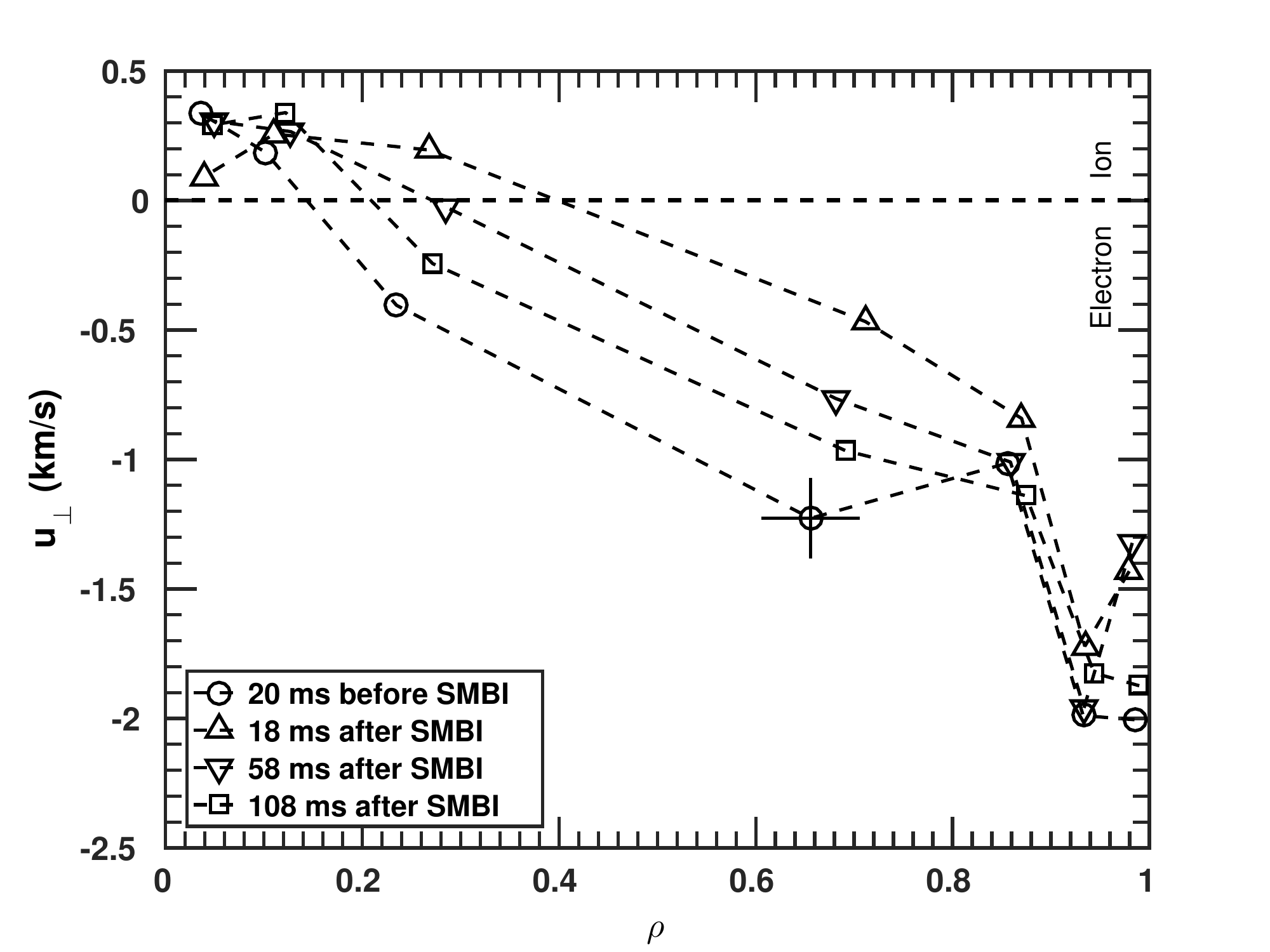}
\includegraphics[width=6cm]{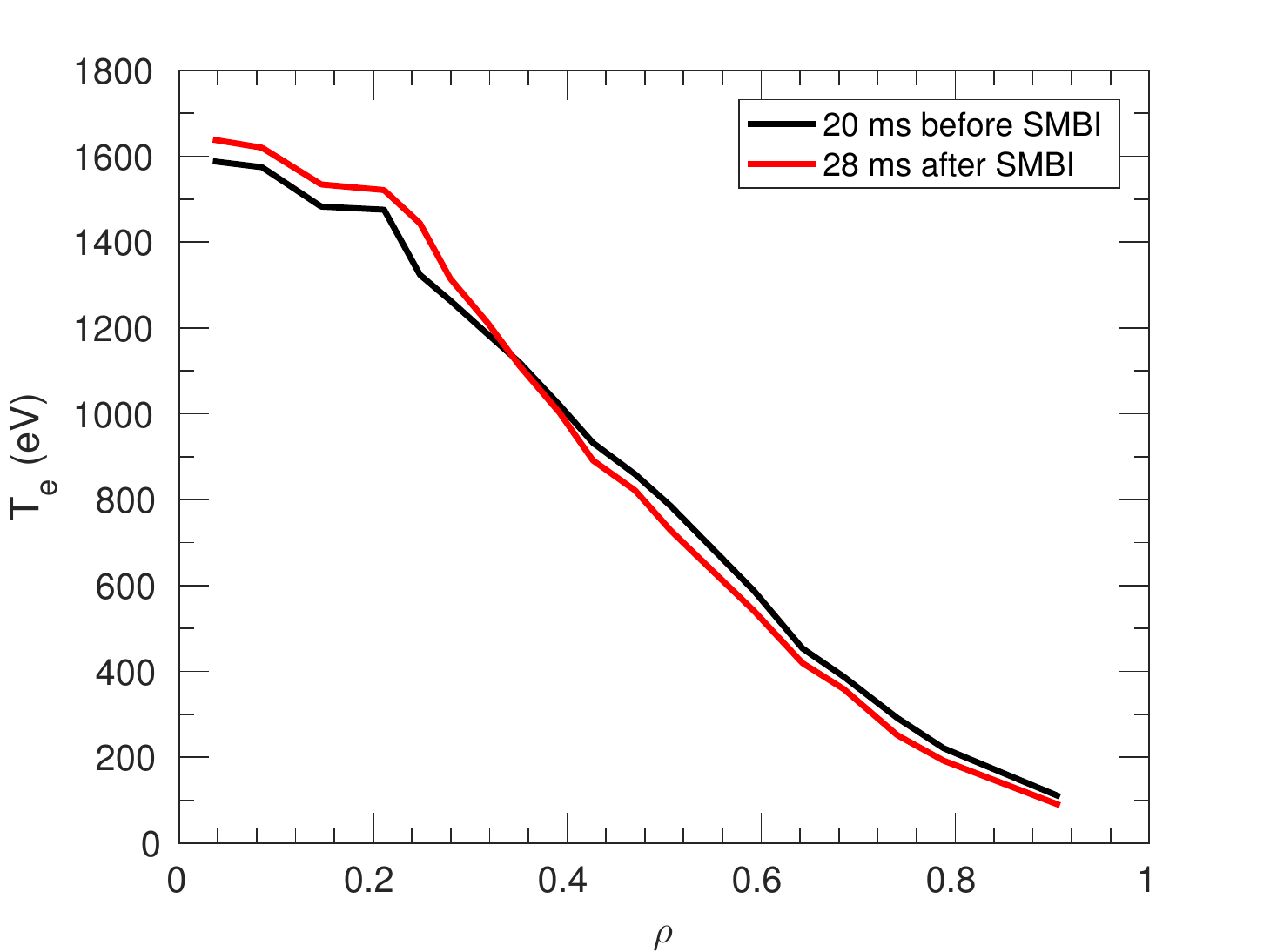}
\end{figure}

The radial profiles of the turbulence perpendicular velocity $u_\perp$ extracted from the Doppler shift measured by DBS system are shown in Figure \ref{fig:u_perp75609}. It is obvious that $u_\perp$ decreases in the radial range $\rho \approx$0.2-0.7 and at the plasma edge. In principle, $u_\perp$ is the sum of the plasma flow velocity $v_{E\times B}$ in the lab frame and the phase velocity of the density fluctuation $v_{phase}$ in the plasma frame: $u_\perp=v_{E\times B}+v_{phase}$. It has been demonstrated \cite{conway2004} that $v_{phase}$  is much smaller than $v_{E\times B}$ in the plasma edge. Therefore, the results indicate that the plasma flow velocity and its shear kept almost the same in the radial range $\rho \approx$0.85-0.95. The reduction of $u_\perp$ at $\rho \approx$1 could be caused by cooling effects of SMBI which increase the damping rate of the plasma flows. It is interesting that $u_\perp$ changes sign after SMBI at the radial position $\rho \approx$0.3, and the reversed polarity responses of the electron temperature occur at a similar radial position. In the plasma core, $v_{phase}$ is comparable with $v_{E\times B}$. From the radial force balance equation, $v_{E\times B}$ is comprised of three components: $v_{E\times B} = v_\phi\frac{B_\theta}{B} - v_\theta\frac{B_\phi}{B} - \frac{\nabla P}{qnB}$. However, it is impossible to evaluate the contribution of $v_{E\times B}$ and $v_{phase}$ due to the lack of plasma rotation measurement with high spatial and temporal resolutions at the moment. 

%\begin{figure}[htbp]
%\centering
%\caption{\label{fig:time_response75609}Evolution of electron temperature, electron density, normalized electron density gradient, and electron density fluctuation at a similar location.}
%\includegraphics[width=8cm]{time_response.pdf}
%\end{figure}

In summary, non-local heat transport has been observed and studied in EAST ohmic plasmas. Both electron temperature and density in plasma core begin to increase roughly 10 ms after SMBI. The core electron temperature reaches its maximum almost at the same time when the edge electron temperature reaches its minimum, and it is roughly 30 ms after SMBI. Inwardly propagation of electron density perturbation induced by SMBI is observed, and this results in a drop of the normalized electron density gradient. The evolution of normalized electron density gradient is of a similar timescale with that of electron temperature in the plasma core. In contrary to the response of core electron temperature and density, the electron density fluctuation increases promptly in the plasma core after SMBI. The amplitude of electron density fluctuation reaches its maximum around 15 ms after SMBI, and drops to a similar amplitude before SMBI within roughly 50 ms. Even though recent simulation work \cite{fernandez2018} agrees well with the cold-pulse dynamics at Alcator C-Mod, our results indicate that the proposed theory may be incomplete in explaining the cold-pulse phenomena in tokamak plasmas in general. As pointed out in previous studies, turbulence spreading may play an import role. The direct observation of electron density fluctuation level increase coincides with the results from the turbulence spreading model \cite{hariri2016} and can be seen as an indicator that the turbulence level raises quickly after the SMBI event. Whereas, it is still unclear and not discussed in the present work how the turbulence enhancement is connected with performance improvement of the electron heat transport in the plasma core. For further investigations it is necessary to investigate several transport channels simultaneously and with good spatial and temporal resolutions. A systematic scan at EAST towards the critical density, revisiting the LOC-SOC transition under the aspect of cold pulse propagation will be proposed. 

\ack{}{}
This work was supported by the Innovative Program of Development Foundation of Hefei Center for Physical Science and Technology, and the National Natural Science Foundation of China under Grant No. 11405211. This work was also partly supported by National Magnetic Confinement Fusion Science Program of China under Contract Nos. 2015GB101003 and 2015GB103002, Basic Science Research Program through the National Research Foundation (NRF) funded by the Ministry of Science and ICT of Republic of Korea (No. 2018R1A2B2008692).

%\newpage
\bibliographystyle{iopart-num}

\bibliography{reference}

\end{document}